# Boundary conditions and the dynamics
# of a dissipative granular gas: slightly dense case

## P. Evesque
### Lab MSSMat, UMR 8579 CNRS, Ecole Centrale Paris
### 92295 CHATENAY-MALABRY, France, *e-mail*: pierre.evesque@ecp.fr

**Abstract:**

*The effect of different possible kinds of motion of the exciting walls (cyclic, random, ...) is investigated on the dynamics of a granular dissipative gas. It is shown that the real distribution of speed of the wall which interact with the balls depend strongly on the real ball speed, due to a screening effect, so that the transfer of excitation from the walls to the cloud of particles may drastically depend on the ball speed itself. This makes the excitation quite non linear, with a change of efficiency law. This may explain the observation of a biphasic cloud when the mean free path of balls is slightly smaller than the cell size L, or may predict that the particles in the cloud is merely at rest as soon as the mean free path is smaller than L/10. It is obvious that the vibrating cell cannot be considered as a thermostat.*

## 1- Status of the problem:

Recently, much work has been devoted to granular gas [1-20]; and hydrodynamics models have been applied which seem to fit simulations [1-5]. In the present work, the problem is revisited in a very simple manner: the paper investigates the true effect of boundary conditions on the statistics of particle speeds in a gas, when the boundary motion is described precisely. It will be demonstrated that the efficiency of the coupling between the particles of the "gas" and the boundary depends strongly not only on the boundary speed distribution, but also on the precise motion of the boundary compared to the true ball speed so that long term correlation can merge and hide the main part of the boundary speed distribution. The notion of temperature is then revisited. These effects are shown to be quite important and to dominate the coupling in most experimental situations when the motion law Z(t) of the boundary is periodic and not random. This occurs when dissipation in granular gas becomes quite large due to grain-grain collisions, *i.e.* as soon as the system overpasses the Knudsen regime, so that the typical excitation speed becomes faster than the typical speed of the balls forming the gas.

     In order to exemplify the contradiction, let us consider simply the wall as a thermostat. In thermodynamics, a thermostat shall be massive so that its mass M (and its heat capacity) shall be much larger than the ball mass m (& ball heat capacity) ; hence the typical standard deviation of speed <V²> of the wall shall be much smaller than the one (<v²>) of the balls (gas), since the two temperature are assumed to be equal, one gets ½ m<v²> = ½ M<V²>, which reads <v²> = <V²>(M/m).

On the other hand, when dealing with granular gas, one expects (or assumes) that the speed of the ball be related to the typical speed of the walls and one writes often that $<v^2>$ scales as $<V^2>$, or $<v^2> = <V^2> f(N_{balls})$, where $N_{balls}$ is the number of balls forming the gas in the container (f is some function). As a matter of fact, it results that the problem stated as above looks rather in contradiction with the first approach, *i.e.* $<v^2> = <V^2>(M/m)$. So few questions arise: can one define a correct statistics of the excitation? Is the temperature a correct notion in the case of a dissipative gas [11]? ....

This raises also the problem of the quantities of physical interest which have to be considered at local and larger (mesoscopic and macroscopic) scales. Are they the local speed v and its standard deviation $<v^2>$. But in this case how can one include the physics of collisions, and the global rules of the mechanics at a mesoscopic scale; and how can one include scaling laws through coarse-graining and averaging? So they shall be rather m**v** and $mv^2$. But in this case, coarse-graining should be efficient everywhere and the concept of temperature should also apply efficiently for the boundary, leading to the good quantity $\frac{1}{2}M<V^2>$, instead of $<V^2>$. But as M>>m, in any practical case, this would mean in turn that one could agitate a large collection of grains with very little motion of the boundary, which is a stupid unphysical hypothesis indeed. So the question remains: why is it ridiculous?

Indeed, one can restate the problem as follows: in classic gas theory, *i.e.* the one for which collisions are not dissipative, the important quantities of physical interest are the particle momentum m**v** and the particle energy $\frac{1}{2}mv^2$, from which one can define mean transfer of momentum and mean kinetic energy. These quantities remain the mechanical quantities of interest at a mesoscopic scale; they are both related to the kinetic tensor $<m(v \otimes v)>$ after local averaging. If such a quantity ($<mv \otimes v>$) is assumed to be independent of the particle mass, as it is found in classic statistical mechanics, then macroscopic average of these quantities becomes linked to the total number of particles in the mesoscopic volume. Local equilibrium of flux of these quantities is ensured in the centre of mass within this hypothesis. Furthermore, local mesoscopic equilibrium is ensured by imposing homogeneous distribution of particles, and local mass flow equilibrium is ensured by assuming homogeneous distribution of each kind of particles.

Still with the same hypothesis of non dissipating collisions, the problem starts being different if the quantities of interest at some point of the system become the speed v and its standard deviation $<v^2>$ instead of the momentum mv and kinetic energy $\frac{1}{2} mv^2$, as it (seems to) occur(s) near a vibrating boundary. Let us imagine two kinds of particles with different masses $m_1$ and $m_2$ ($m_1 > m_2$); then far from the boundary, local collisions shall ensure redistribution of energy and momentum among the particles, enforcing a macroscopic equilibrium of flux, and the equi-distribution of kinetic energy. It results from this that the typical speed of particle 1 (the heavier one) shall be smaller than the one of particle 2; such a situation shall occur as soon as bulk condition has started to be encountered. Hence, under such a hypothesis, different particles with different masses shall behave differently near the boundary and deeper into the pile even when collision between particles preserve energy; then they shall not have the same statistics, and this statistics shall vary with the distance to boundary [3,4].

In particular, such a situation shall be revealed when particles come back towards the boundary to gain energy. The difference of speed between the two kinds of particles may result in a difference of speed distribution after the collision with the boundary, leading in fact in a different condition from $v^2=V^2$ for both of them, as stated initially, and leading also to some kind of spontaneous segregation. So what are the real boundary conditions which have to be assumed when dealing with dissipative granular gas?

This is just the topic of this paper. It will be answered by investigating the real dependence of the typical speed $v'_+$ of a ball after the collision with the boundary on its typical speed $v_-$ before the collision, taking into account the speed V of the boundary and the probability of the collision.

One can note for instance that simulation works and theoretical descriptions based on kinetic theory of particles use often $\langle v^2 \rangle = \langle V^2 \rangle$ as correct boundary conditions, whatever the excitation shape is supposed to be, indifferently whenever it corresponds to a true thermal excitation, or to a sinusoid motion of the boundary, or to a saw tooth excitation. In the present paper, on the contrary, *we show that some important difference is generated by these few types of excitation*.

It is now known from experimental results in micro-gravity condition, which have been concerned by a gas of balls dissipating due to collisions and excited by a sinusoid motion of the container, that the speed distribution of this gas is characterised by a typical speed v (or $\langle v^2 \rangle$) smaller than the boundary one V (or $\langle V^2 \rangle$) [7-9]. This has been revealed by the periodic appearance of a gap between the cloud of balls and the container. It results from this that the collisions between the balls and the boundary occur mainly when the boundary is merely at maximum extension towards the gas of particles. As the maximum extension corresponds to V(t)=0, the collisions occurs often (a) when the boundary speed V is small and (b) when it varies fast with time (maximum of acceleration $\Gamma$). Hence the transfer of energy may be quite inefficient.

Furthermore, since the slower the speed of the ball, the more located near the maximum the collision and the smaller the transfer/gain of kinetic energy. It results from this that the excitation process (i) shall depend strongly on the typical mean speed of the gas and (ii) shall be highly heterogeneous depending on the real speed of the ball, since a slower ball will gain much less energy in average than a faster one.

Point (i) makes the excitation coupling efficiency quite non linear as soon as particle-particle collision rule makes the gas dissipative and reduces the typical particle speed ; this occurs as soon as the mean free path of gas particles becomes smaller than the cell length L, or in other word as soon as the gas is out of the Knudsen regime.

Point (ii) tells that energy transfer during the collision with the boundary is more efficient with faster balls than with slower balls. So, if this non linearity turns to be quite efficient, this could lead to consider two different classes of balls, one which is slow and forms a "cold gas" of particles, merely not coupled to the boundaries, but coupled to other faster particles which form the second class; the particles of this second/other class are much faster, since they are much more coupled to the boundary, and they are cooled down by collisions with the "cold gas". This second class would be "reminiscent" in some way of the Knudsen regime at lower concentration. The coupling between fast and slow classes maintains the slow balls in the centre of the cell.

Indeed, this seems to be observed in experiment on granular gas in the Airbus 0g [9-14].

*Comparison with computer works and theoretical approach:*
Anyway, one observes important differences between the experimental behaviour and the behaviour obtained from theoretical approach or simulations in past works, when true boundary conditions are not treated in details. For instance, it has been argued [16] that the experimental results can be modelled with a restitution coefficient $\beta$ which depends on the ball speed v at collision with boundary [16]; but these simulations require using a restitution coefficient which decreases strongly when ball speed becomes large [16]. This is quite unlikely for two reasons: firstly, we have measured the variations of $\beta$ with the speed difference v-V, and have found it is merely constant experimentally till v-V< 2m/s [17]. Secondly it is just the contrary of what predicts the present analysis following point (ii), so that the proposed model does not work correctly.

Other simulations have considered different kinds of excitation, with some law of coupling with the boundary which depends on the speed of the balls [5]. They could lead to statistics similar to the one found in the experiment. But what is the physical meaning of such a hypothesis? Is it the one we present here? In this case it is much better to be told.

However it is really surprising that the true effect of boundary conditions has not been discussed in great details in the past ten years, while the number of published papers concerning granular gas behaviour has grown up largely. For this reason, we predict that it should be quite interesting to compare the full details of simulations of balls moving according to dissipative collisions, and submitted to different motion of boundaries. Even, the ball number could be small, *i.e.* few tens or few hundreds should be enough.

In this paper we do not consider explicitly the role of ball rotation, which increases strongly the dissipation during collision and modifies the inclination of the ball trajectory after the collision. We just want to focus on the effect of time correlation in the boundary motion. A previous paper [10] has been devoted to the study of the boundary conditions for a gas in a Knudsen regime for which very fast particles were detected, whose statistics was scaling anomalously, *i.e.* $\rho(v) \approx (1/v) \exp(-v/v_o)$ ; in this case, particle-particle interaction remains small, even if it is not negligible. The present paper deals more with larger concentration of balls, for which dissipation is large, resulting in balls with extremely slow dynamics. This looks quite surprising at first sight, but it is an experimental fact. And the question which is tentatively answered here is then: can one understand why the balls and the system look so quiet? And the answer proposed here is: it is because boundary conditions of the granular gas depend on the surface-to-volume ratio, so that the coupling to grains tends to zero when this ratio overpasses some value of order 1-10.

*Outline:* The paper is built as follows: Section 2 recalls the classic approach; it defines the statistics of speed distribution of the moving wall and gives few examples of such statistics. However due to some shadow effect, the previous method has to be modified; the basic concept is presented in Section 3, together with the general formulation. Section 4 gives different examples of application to granular dissipative gas submitted to periodic excitation. Section 5 proposes other kinds of excitation based on random motion of the wall. Section 6 of the paper is devoted to a discussion and conclusion.

## 2- Typical speed distribution linked to the motion of boundaries:

Let us first consider the problem of the excitation by a moving boundary. It can be seen first as a classic problem of collision which occurs between two objects in random position. This is this point of view we develop here.

We assume the boundary mass M to be infinite; be m the particle mass; be $V(t)$, $V'(t)$, $v$, $v'$ the boundary speed and the particle speeds before and after the collision respectively; and be $\beta$ the restitution coefficient, the collision rules between ball and boundary writes:

$V'(t)=V(t)$ and $v'-V'=-\beta(v-V)$ or

$v'=(1+\beta)V-\beta v$ (1)

In classic collision approach, as the two objects move randomly compared to each other, the collision probability is just the product of probabilities of finding each object at speed v and V respectively multiplied by the relative volume $(V-v)\,\sigma\,dt$ explored during dt by the piston of surface $\sigma$. So, be $p_b(v)$ and $p_w(V)$ both probabilities, and assuming that $p_b(v)$ is independent of position (distribution uniform in space), the collision probability P becomes:

$P = p_b(v)\, p_w(V)\, |v-V|\, dvdV\, \sigma\, dt$ (2)

This collision produces a particle at speed $v'=(1+\beta)V-\beta v$; and if v was in the range $[v,v+dv]$, $v'$ would be in between $[v'=(1+\beta)V-\beta v, v'-\beta dv]$, so that the indeterminacy $dv'$ on $v'$ due to dv reduces since $dv'= -\beta dv < dv$. This reduction is linked to the dissipative nature of the collision; however, it is counterbalanced by the possibility of V to get different values. To compute the probability $p'_b(v')\, dv'\, dt$ of appearance of v', one has to take into account this reduction on the scale of v' and to renormalize by 1/β, then to sum over all the possibilities of V.

And the probability distribution of appearance per unit of time is p'$_b$(v') = (1/β) ∫$_{such\ as\ v'=(1+β)V-βv}$ p$_b$(v)p$_w$(V)|v-V|dvdV, for which v=[-v'+(1+β)V]/β , which writes also:

$$p'_b(v') = (1/β) \int p_b(\{[1+β]V-v'\}/β) \, p_w(V) \, |v-V| \, dV \quad \text{or equivalently as} \quad (3)$$

$$p'_b(v') = (1/β) \int p_b(v) \, p_w([v'+βv]/[1+β]) \, |v-V| \, dv \quad (4)$$

Hence the new distribution p'$_b$(v') of v' after the collision is known as soon as soon as one knows (i) the distributions of v and of V before the collision, and (ii) one knows the collision rule.

It remains to evaluate p$_w$(V) for different kinetics of wall motion. Be Z(t) the motion law as a function of time; we are interested by cyclic motion, so we assume here that Z(t) is periodic of period T. Be Z'(t)=V(t) and Z''(t)=Γ(t) the speed and the acceleration of the boundary as a function of time, p$_w$(V)dV is the time per period spend by the boundary in between V and V+dV. So as dV=Z'' dt, one has p$_w$(V)dV = dt/T with dt=dV/Z'', and one gets:

$$p_w(V) = |1/(T\, Z'')| \quad (5)$$

where Z'' has to be written as a function of Z'.

***Sinus motion:*** Suppose Z' to be sinusoid, then Z'(t)= −aω sin(ωt), or t = (1/ω) Arcsin{-Z'/(aω)}; also 1/(T Z'')= −1/[aTω² cos(ωt)] and

$$p_w(V)= 1/[2πaω \cos(\text{Arcsin}\{-Z'/(aω)\})]. \quad \text{For sinus motion.} \quad (6)$$

We give in Fig. 1.a this function. (Writing Z''(t)=-aω²cos(ωt)=ω(a²ω²-Z'²)$^{½}$ in this sinus case, one gets equivalently p$_w$(V) δV= δV/[2π(V$_o$²-V²)$^{½}$] for Eq. 6).

So, a maximum of |V| exists, which corresponds to |V|=aω, and the distribution tends to infinity when |V| is at maximum. This is obvious since it corresponds to Z''=0 there. The distribution is minimum at V=0 which corresponds to maximum acceleration. The distribution is symmetric compared to vertical axis in this peculiar case. This depends on the symmetry of Z(t).

***Archs of parabola:*** Fig. 1.b report the distribution obtained with excitation imposing constant +Γ & -Γ accelerations during half periods alternatively.

***Saw tooth motion:*** In the case of a saw tooth motion with two speeds V$_1$>0 and V$_2$<0 during two lapses of time τ$_1$ and τ$_2$, such as τ$_1$V$_1$ + τ$_2$V$_2$ =0 and τ$_1$+ τ$_2$=T, the distribution should be composed of two Dirac distribution located at V$_1$ and V$_2$ only, with respective weight $T/\tau_1$ and $T/\tau_2$, as in Fig. 1c. However, periods of constant acceleration/deceleration have to be introduced to let the system obeying the law of mechanics, so that a more realistic probability distribution should look as in Fig. 1.c, which includes some horizontal straight line at some non zero height.

*Gaussian distribution:* A Gaussian distribution of speed can be obtained if the system is submitted to a viscous drag with a time constant $T_R=1/\mu$; one needs just to impose a random series of fixed constant acceleration $\pm\Gamma$ lasting a short constant lapse of time $\delta t$ before changing randomly of sign($\pm\Gamma$) or direction, with $\delta t \ll T_R$. In this case the speed overcomes a random walk with step $\delta V=(\pm\Gamma)\delta t$; so, the mean speed of the system is 0, its quadratic speed $<V^2>$ increases linearly with time as $<V^2>=\delta V\, t$, till it saturates after a time $t= T_R$ to reach a constant value $<V^2> = \Gamma^2\, T_R\, \delta t$. If the viscous drag is not introduced, the mean square displacement will increase linearly to infinity (see Appendix).

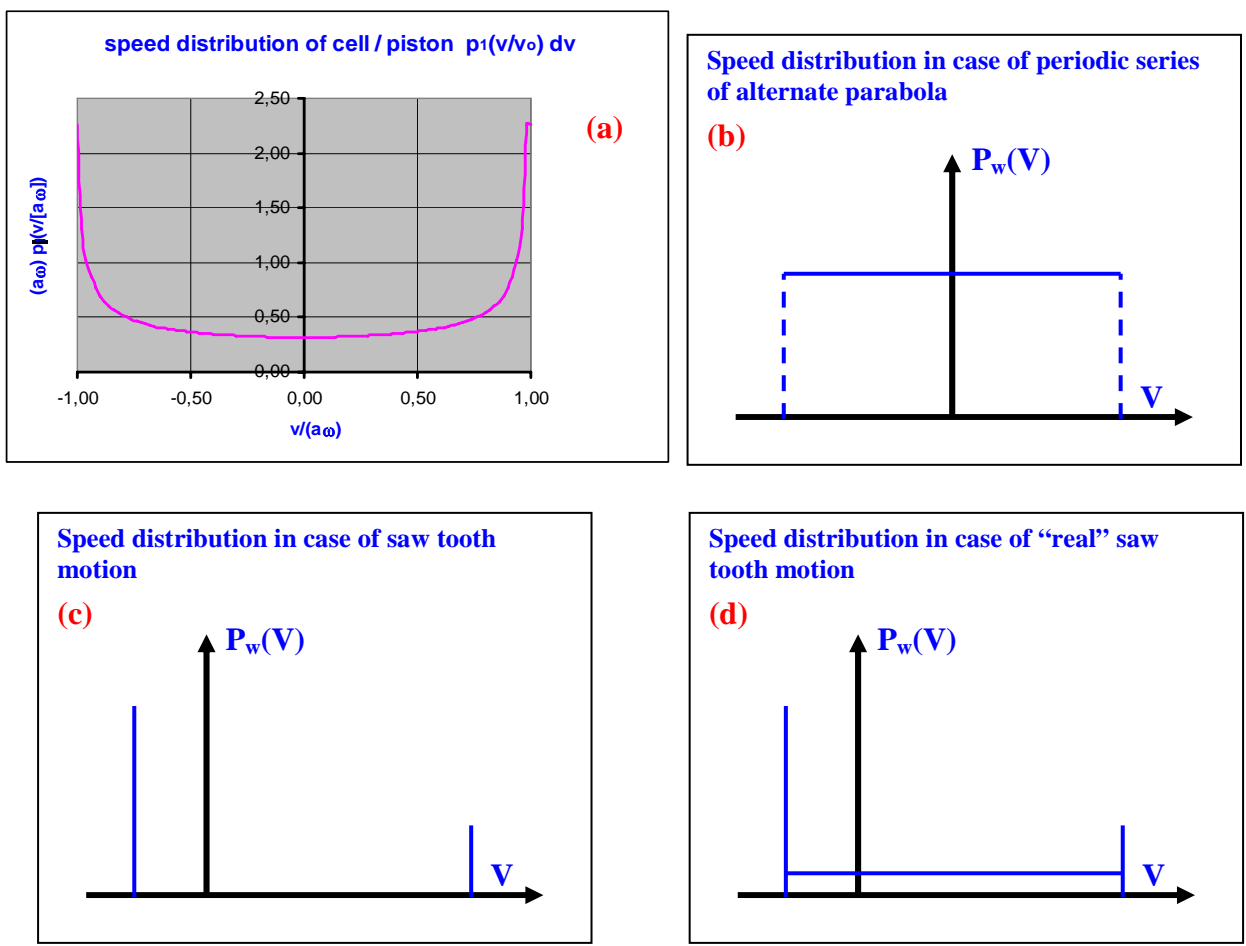

*Figure 1:* Probability distribution of the piston speed (a) when the piston is moving sinusoidally, period $T=\omega/(2\pi)$, amplitude a, (b) according to two arches of parabola obtained by submitting the system to constant acceleration $\Gamma$ and $-\Gamma$ during half a period each, (c) according to a theoretical saw tooth law with two speeds $V_1$ and $V_2$., (d) according to a real saw tooth law which takes into account periods of constant acceleration / deceleration .

In the case of cyclic motion of the wall, the wall passes at least twice per period at the same position, so that it gets at least twice the same speed, once with Γ>0, the other one with Γ<0; but these two Γ may be different so that the distribution of speed shall be computed for both directions and summed then, in the general case. This is not necessary for symmetric motion such as a sinusoid, for which each way gives the same distribution.

### 3- Real Typical gain of speed after a collision.

Since we know now the distribution of speed of the boundary, we may want to know the effect of the collision on the ball speed statistics. This means to use *a priori* the previous modelling, *i.e.* the $p_w(V)$ determined in the last section combined with Eqs. (1-6). In fact this cannot be done as simply, because one has to take into account some trajectory geometrical constraints which are exemplified in Fig. 2, where we have drawn the Z position of the boundary and the z one of the ball all together.

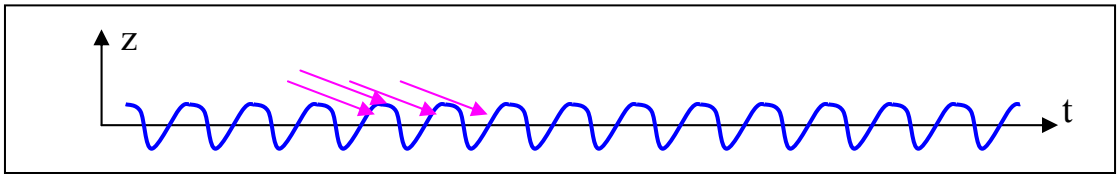

***Figure 2:*** *in blue the position of the piston obeying a periodic motion: Z = Z(t)=f(t) => V = Z' (t) and dV/dt=Z"(t). In pink the trajectory of different balls colliding with the piston.*

A first constraint is that the ball speed v shall be always smaller than the wall speed V, since the balls are located on top of the wall. This reads v-V<0, or V>v, (v is negative). This makes a first threshold, which exists only when the speed v of the ball is slower than the maximum speed $V_{max}$ of the wall, *i.e.* $|v|<|V_{max}|$. It cuts the V distribution in two regions and forbids the one with the main negative part.

Such exclusion, *i.e.* V(t)<v, exists also in model of section 2. However, the forbidden zone extends further here: the cut extends and hides some part of the distribution for which v is smaller than V(t) (v is negative). This is because some points of the trajectory cannot be reached by the balls due to some shadow effect (see Fig. 2). This second effect shall appear as soon as v<V, (with v<0); but as long as $|v/V_{max}|$ is not too small, the shadowing remains essentially concerned by the part of the Z trajectory located near the minimum and it forbids only the part of the V distribution which is concerned either by very negative values or slightly positive ones. Hence this effect improves the coupling between the boundary and the wall when $|v|/V_{max}$ is not too small. On the other hand, when |v| becomes quite smaller, this effect starts hiding some part of the fastest positive values of V. If T is the period of the motion, this occurs approximately when $|v|T<Z_{max}/2$ about, with $Z_{max}$ being the amplitude of the wall motion (the true value depends on the exact Z(t)).

### *3.a- General calculation:*

In this part we calculate first the shadow zone. Then we consider the case of fast ball for which there is no shadowing effect ($v_b<V$) (or $|v_b|>V_{max}$). Then we calculate the system with a shadow.

### *Shadow zone:*

Be f(t) the position of the wall with time, $V(t)=f'(t)$ and $\Gamma(t)=f''(t)$ its speed and acceleration respectively; the wall motion is periodic of period T, with $f(0)=Z_{max}$ its amplitude (since the mean position is assumed equal to 0). We will refer the trajectory of the balls by their speed $v_b$ and by the time t' at which they cross the horizontal line passing at height $z=z_{max}$. Hence this line joins the maxima of f(t) (see Fig. 3). To find the hidden part of the trajectory, we need to define the critical trajectory of the ball; this one corresponds to the trajectory of the ball which is tangent to Z(t) in its highest part (at $T_1$) and which crosses Z(t) some time later at $T_2$, for the first time; hence :

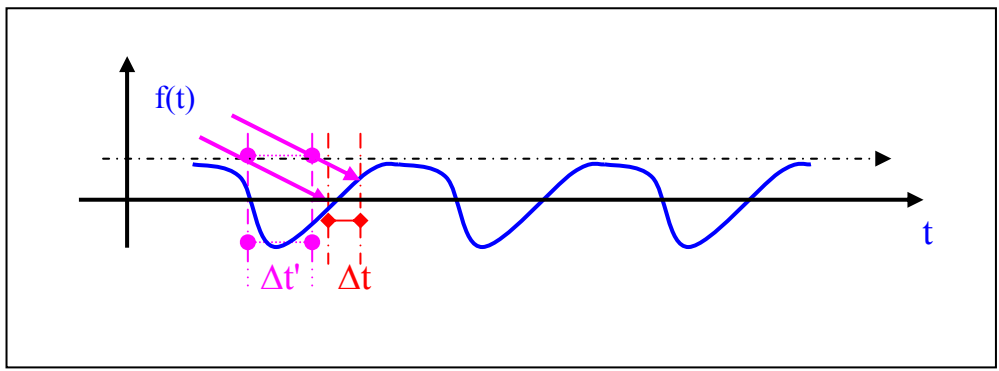

*Figure 3 :* Definition of t', t, v: f(t) is the piston position, $Z_{max}$ its maximum height. 2 balls collide the piston at two locations $z_1=f(t_1)$ and $z_2=f(t_2)$, very near so as $V_2=V_1+\delta V$; these two balls were crossing $Z=Z_{max}$ at $t'_1$ and $t'_2$, such as $f(t)=0+v_b(t-t')$

$0<T_1<T, 0<T_2-T_1<T$ and  (7.a)

$f'(T_1)=v_b$  (7.b)

$T'_1=T_1+[z_{max}-f(T_1)]/v_b =T_2+[z_{max}-f(T_2)]/v_b$.  (7.c)

There are two values of t per period which satisfy Eq. (7.b); $T_1$ corresponds to the shortest time after the maximum of Z. The points of the wall trajectory which are not allowed for collisions correspond to Z values below this trajectory of ball, at each period. Collisions occur in between intervals $[T'_2+nT, T'_1+(n+1)T]$; (but each of these periods correspond to trajectories passing at $Z_{max}$ in between $[T'_1+nT, T'_1+(n+1)T]$ so that all trajectories are colliding).

### *3.b  Speed distribution when no shadowing:*

We consider first the case when $f'(t) > v_b$ ($v_b$ is negative). It means that any Z position of the piston is reachable by a ball (no screening). We consider a constant flow of balls at speed $v_b$, which means that the number of balls N per unit of time which crosses the $Z=Z_{max}$ line is uniform. We consider two balls hitting the wall at time $t_1$ and $t_2 > t_1$, such as $t_2-t_1=\delta t$ is small. They have crossed the $Z=Z_{max}$ line at time $t'_1$ and $t'_2$ respectively, with $\delta t'=t'_2-t'_1$; they collide the piston at time $t_1$ and $t_2$ such as: $f(t)=Z_{max}+v_b(t-t')$ so that one has for both collisions:

$$f(t_1)= Z_{max}+v_b(t_1-t'_1) \quad \text{and} \quad f(t_2)= Z_{max}+v_b(t_2-t'_2) \quad \text{so that}$$

$$f(t_2)- f(t_1) = v_b(t_2-t'_2 -t_1+t'_1)= v_b(\delta t-\delta t') \tag{8}$$

On the other hand, expanding $f(t_2)- f(t_1)$ one gets:

$$f(t_2)- f(t_1) \approx f'(t_1)\, \delta t + f''(t_1)\delta t^2/2 +\ldots= V(t_1)\, \delta t + \Gamma(t_1)\delta t^2/2 +\ldots \tag{9}$$

Then equating Eqs. (8) and (9) leads

$$[v_b-V(t_1)]\, \delta t = v_b\delta t' + f''(t_1)\delta t^2/2 +\ldots \tag{10}$$

In the same time, one can write that the difference $\delta V$ of speed of the piston in between $t_1$ and $t_2$ is $\delta V=f''(t_1)\delta t=\Gamma(t_1)\delta t$ at first order, so that one gets at to first order

$$\delta t'= [\delta V/ f''(t)]\, (v_b-V)/v_b \tag{11}$$

The number of balls per unit of time which are at speed $v_b$ and which hit the piston at speed V can be written as $p_w(V)\, \delta V$. It is given by $p_w(V)\, \delta V = N\, \delta t'/T$, with $\delta t'$ obeying Eq. (11). Hence $p_w(V)$ is:

$$p_w(V)= N\, (v_b-V)/[T\, v_b\, f''(t)] \tag{12}$$

   with $f'(t)=V$ and $\Gamma=f''(t)$, $\Gamma\delta t=\delta V$

   This Eq. (12) is equivalent to Eq. (2).
   So if $f''(t)$ can be 0, it corresponds to a maximum of density of collision. This occurs when V is a maximum and when V is opposite to $v_b$. When $p_w(V)$ is always positive, whatever V, the piston can be touched at any time. When the motion is not symmetric, determination of $p_w(V)$ has to be performed for both the accelerating zone and for the decelerating zone of the piston trajectory.

If $v_b-V$ is negative on some part of the trajectory, the collision can occur with some ball on the other side of the wall only, hence with a ball outside the system. So this solution corresponds to unphysical condition and has to be omitted. This part of piston trajectory is not reachable during that lapse of time, but we have seen also that this part of the trajectory generates some shadow on some adjacent part of the piston trajectory. This last one is not reachable too. So limits have to be computed in the (Z, V, Γ) space using the values of $T_1$ and $T_2$ calculated from the set of Eq. (7).

Eq. (12) is useful mainly when one knows the acceleration Γ of the piston as a function of its speed, Γ(V). This is the case for sinus motion for instance.

$$p_w(V) \, dV = N \, dV \, (v_b-V)/[T \, v_b \, \Gamma(V)] \quad (13)$$

And the speed $v'_b$ after the bouncing is given by Eq. (1), *i.e.* $v'_b = (1+\beta)V - \beta v_b$. In this case, average speed can be computed by integration over all possible values of V… and mean field theory can be developed.

However, the problem of the integration domain remains to be determined, which can only be done analysing the crossing of the trajectories and using $T_1, T_2, T'_1$ and $T'_2$. The way to proceed will be exemplified in the next section

## 4- Application to granular gas

In experiments on granular gas in micro-gravity, it has been found that a depletion zone exists near the boundary. It has been interpreted has linked to the fact that balls are moving slower than the boundary. This result holds true as soon as the mean free path $l_c$ of the ball in the "gas" is larger or of the order of the cell length L. And the typical speed $<v_b>$ decreases fast when $l_c$ is decreased. However it has been found that $<v_b>$ is much larger than $V_o = a\omega$ in the case of a very rare number of balls (*i.e.* 1 or 2 balls).

As already told, and as can be seen on Figs. 1 and 2, shadowing cuts some part of the V distribution which is no more accessible; when $v_b$ starts being too small, this cut starts at $V_1 = v_b$ and goes towards negative V. If $|v_b|$ is not too small the domain stops at some negative $V_2 < 0$; this has not much effect except to improve the coupling of the piston with the balls since coupling occurs essentially when the piston goes fast toward the balls. Hence this shadowing starts by improving the efficiency of the coupling.

However, for smaller values of $|v_b|$, one sees that small positive values of V start also to be forbidden. Finally when $v_b$ becomes even smaller, the larger values of V start also to be forbidden, so that the coupling starts to become quite little. This effect starts becoming important when $v_b T << a = Z_{max}$, the amplitude of wall motion. Such condition is encountered experimentally in granular gas whose mean free path is some tenth of the cell length L, since $v_b/V_{max}$ has been found to be quite small in this case. So the effect of this phenomenon cannot be neglected anymore.

In other words, shadowing inhibits the coupling when the particles can only hit the piston near the maximum of piston position, *i.e.* when the piston speed is quite slow at this phase of the period. It is worth noting that this effect will occur whatever the kind of excitation shape, because inertia of the system will force the trajectory of the piston to be curved near its extrema. This is obviously true for a sinus shape or for a shape consisting in an alternate series of positive and negative arches of parabola; but this occurs also for a saw tooth shape, when the balls are able only to hit the upper part of the teeth for which inertia will make the tooth round.

As we want to quantify the effect of the shadowing, we will need also to compare the results to the case when no shadow exists. We first consider the sinus motion; then we extend the result to a trajectory corresponding to two arches of parabola oriented in opposite direction. We will end with consideration on sawtooth motion.

### *4.a- case of sinus motion*

#### *Sinus motion with important shadowing:*

Be $Z = a \cos(\omega t)$ the equation of motion of the wall. We consider only the distribution for the sinus motion, when there are two main trends obtained in the two limit cases: the first one is such that $v_b \gg V_o = a\omega$, the second one is when $v_b \ll V_o$. Furthermore, one sees that the distribution of V is rather flat as soon as one considers speeds V smaller than $V_o/1.5$ or $V_o/2$, so that it can merely be approximated by a constant $\Gamma = a\omega^2$. Hence, this corresponds to consider the wall motion mainly governed by some constant acceleration $\Gamma$.

The domain of collision speed is limited by $V_1 = v_b$ on one side and by some speed $V_2 > 0$. A way to compute $V_2$ is to remember that the integral $\int_{v_b}^{V_2} p_w(V) dV$ shall be equal to the ball flux N; so, one gets $N = \int_{v_b}^{V_2} N \, dV \, (v_b - V)/[T \, v_b \, \Gamma(V)] = N(V_2 - v_b)^2/(2Ta\omega^2 v_b)$.

$$V_2 \approx v_b + (4\pi a\omega v_b)^{1/2} \qquad (14)$$

We can also calculate the average speed after the bounce. It is given by :

$$\langle v'_b \rangle = \int_{v_b}^{V_2} v'_b \, dV \, (v_b - V)/[T \, v_b \, \Gamma(V)] \qquad (15)$$

With $v'_b = (1+\beta)V - \beta v_b = (1+\beta)(V - v_b) + v_b$. Hence

$$\langle v'_b \rangle = \int_{v_b}^{V_2} dV \, (1+\beta)(V-v_b)^2/[2\pi a\omega \, v_b] + \int_{v_b}^{V_2} dV \, (V-v_b)/[2\pi a\omega] \qquad (16)$$

$$\langle v'_b \rangle = v_b + (2/3)(1+\beta)(4\pi a\omega v_b)^{1/2} \qquad (17)$$

We can assume now that the action of neighbouring gas will be to slow down the speed of this ball by a coefficient $\alpha \ll 1$ before the ball comes back to hit the wall. Hence it will come back on the wall with the mean speed $\langle v''_b \rangle = -\alpha \langle v'_b \rangle$. And a self consistent hypothesis applied on the mean speed tells $\langle v''_b \rangle = v_b$ and leads to:

$$\langle v_b \rangle = \alpha v_b + (2\alpha/3)(1+\beta)(4\pi a \omega v_b)^{1/2} \tag{18}$$

$$\langle v_b \rangle \approx [\alpha(1+\beta)/(1-\alpha)]^2 (16\pi/9) a\omega \approx 20\, \alpha^2\, a\omega \tag{19}$$

In Eq. (19) we have used the fact that $\beta \approx 1$ and $\alpha \ll 1$. The calculation has been performed assuming that $v_b T \ll a$; this writes $2\pi v_b \ll a\omega = V_o$. Combining this condition with Eq. (9) leads to $120\alpha^2 \ll 1$, which gives $\alpha < 0.05$ as the domain of validity.

*Sinus motion with no shadowing:*

According to previous definition of times t and t', we know that the ball density is uniformly distributed according to t', but that the ball hits the boundary at position z slightly later such as $z = z_{max} + v_b(t-t') = z \cos(\omega t)$. This leads to $v_b dt' = (v_b - V) dt$, with $V = -a\omega \sin(\omega t)$. After the collision, the ball comes back with the speed $v'_b = (1+\beta)V - \beta v_b$. So, the average speed $\langle v'_b \rangle$ after the collision is given by:

$$\langle v'_b \rangle = \int_0^T [(1+\beta)V - \beta v_b](v_b - V)\, dt \,/\, \int_0^T (v_b - V) dt \tag{20}$$

$$\langle v'_b \rangle = \int_0^T [-(1+\beta)a\omega \sin(\omega t) - \beta v_b](v_b + a\omega \sin(\omega t))\, dt \,/\, \int_0^T (v_b + a\omega \sin(\omega t)) dt$$

$$\langle v'_b \rangle = -\beta v_b - (1+\beta)(a\omega)^2/(2v_b) \tag{21}$$

After the collision the particle propagates across the gas and comes back with a mean speed $\langle v''_b \rangle \approx -\alpha \langle v'_b \rangle$. Following the standard mean field approach, with some self consistent condition, one writes $\langle v''_b \rangle = \langle v_b \rangle$:

$$\langle v_b \rangle = \alpha\beta v_b + \alpha(1+\beta)(a\omega)^2/(2v_b) \tag{22}$$

Indeed this is possible because the term $v_b$ is supposed to be large compared to $a\omega$; hence the term $(1+\beta)(a\omega)^2/(2v_b)$ is supposed to be smaller than the term $\beta v_b$.

Let us first consider the case $\alpha=1$, which corresponds to a single ball case. One gets:

$$\langle v_b \rangle = -a\omega \{(1+\beta)/[2(1-\beta)]\}^{1/2} \quad \text{(single ball case, } \alpha=1\text{)} \tag{23}$$

Same result has been obtained in ref [18]. Since $1-\beta$ is small quite often, this leads to the condition $v_b \gg a\omega$. This is a known result. Also, in the limit of no dissipation the typical ball speed becomes infinite, because its probability to get energy during the bouncing shall be equal to the probability of loosing energy. So the ball "temperature" shall be infinite, compared to the boundary- speed/temperature. This makes the boundaries behaviour working quite differently from a classic thermostat.

However, coming back to the granular gas case, with $\alpha<1$, one gets from Eq.(22) as soon as $\alpha$ is small:

$$<v_b> = -(a\omega) [\alpha(1+\beta)]^{1/2}/[2 -2\alpha\beta)]^{1/2} \approx [\alpha(1+\beta)/2]^{1/2} a\omega \qquad (24)$$

Also Eq. (24) indicates that $<v_b>$ becomes smaller than $a\omega$ for $\alpha<0.5$ when $\beta=1$. Also, $<v_b>$ scales linearly with $V_o=a\omega$. Comparison of Eq. (19) and (24) shows the effect of shadowing which transform the $\alpha^{1/2}$ dependence into the $\alpha^2$ dependence. This demonstrates the importance of shadowing for the reduction of coupling. Validity of Eq. (19) requires $\alpha<0.05$ as studied earlier, Crossing occurs for $0.2<\alpha<0.3$, .

## *4.b- Case of two arches of parabola:*

If the motion is made of two arches of parabola oriented alternately in opposite directions, $\Gamma$ would be just the acceleration, and the previous calculation is correct. Be T the period, $Z_o$ the maximum height $Z_o= \frac{1}{2}\Gamma(T/4)^2$, and the maximum speed $V_o=\Gamma T/4$. The parabola equations write:

$$Z=Z_o-1/2\Gamma(t-nT)^2 \qquad t\in [-T/4+nT, T/4+nT] \qquad (25.a)$$
$$Z= -Z_o+1/2\Gamma(t-nT-T/2)^2 \qquad t\in [T/4+nT, 3T/4+nT] \qquad (25.b)$$

*Important shadowing:* When shadowing exists and when it is large, one of the limit of the shadowing is $T_1= -v_b/\Gamma$, with $Z_1=Z_o-v^2_b/(2\Gamma)$; when $v_bT<Z_o$, the second limit $T_2$ is solution of :

$$Z=Z_1+v_b(T_2-T_1)=Z_o-\Gamma(T_2-T)^2/2= Z_o-v^2_b/(2\Gamma)+v_b(T_2-T_1) = Z_o+v^2_b/(2\Gamma)+v_bT_2$$

$$0= \Gamma(T_2-T)^2/2 + v_bT_2 + v^2_b/(2\Gamma) \qquad (26)$$

But since the distribution of piston speed is uniform on V, the best way to proceed is to use the same method as proposed for sinus motion, and tells that the integral of $p_w(V)dV$ within the limit of integration shall be equal to N; hence $N= \int_{v_b}^{V_2} N \, dV \, (v_b-V)/[T \, v_b \, \Gamma] = N(V_2-v_b)^2/(2T\Gamma v_b)$; or $V_2= v_b +(2T\Gamma v_b)^{1/2}$. Similarly, one can compute the typical speed $<v'_b>$ after the bouncing for a given $v_b$, using Eq. (15)

$$<v'_b> = \int_{v_b}^{V_2} v'_b \, dV \, (v_b-V)/[T \, v_b \, \Gamma] \qquad (27)$$

With v'$_b$=(1+β)V-βv$_b$= (1+β)(V-v$_b$)+v$_b$ . Hence

$$<v'_b> = (1+\beta)(V_2-v_b)^3 / [3T\, v_b\, \Gamma] + v_b = v_b + (2/3)(1+\beta)(2T\Gamma v_b)^{1/2} \quad (28)$$

Eq. (28) is similar to Eq. (17).

***no-shadowing case:*** To estimate $<v'_b>$ when there is no shadowing, it is better to use the time integral directly, as used previously (Eq. 20).

$$<v'_b> = \int_0^T [(1+\beta)V-\beta v_b](v_b-V)\, dt \,/\, \int_0^T (v_b-V)dt \quad (29)$$

As (i) the integral on a period of V is 0, (ii) since $V_o=\Gamma T/4$, (iii) since V varies linearly with t from $-V_o$ to $V_o$, (iv) since $\int_{T/4}^{5T/4}(v_b-V)dt = v_b T$ and $\int_0^T [(1+\beta)V-\beta v_b](v_b-V)\, dt = (1+\beta)\int_0^T V^2 dt - \beta v_b^2 \int_0^T dt = -(1+\beta)\Gamma^2(T)^3/48 - \beta v_b^2 T = (1/3)(1+\beta)(V^2_o)T - \beta v_b^2 T$ , one gets:

$$<v'_b> = -(1+\beta) V^2_o/(3v_b) - \beta v_b \quad (30)$$

Both terms on the right hand part are positive since v$_b$ is negative. This trend compares quite well with the one found for sinus shape (Eq. 21) . Similar treatment can be achieved to determine $<v_b>$, which assumes a mean field behaviour with loss α due to the propagation of the particle in the dissipative gas and with self-consistent hypothesis, leading to a trend similar to the sinus case, and a similar discussion: when α=1 (1 ball case), the speed $<v_b>$ becomes quite large, tending to infinite when β tends to 1, while it reduces strongly as soon as α<0.2….

### *4.c- Case of a sawtooth motion:*

If the motion is sawtooth-like, characterised by two speeds $V_1>0$ and $V_2<0$ , lasting respectively $\tau_1$ and $\tau_2$ so that $\tau_1+\tau_2=T$ and $\tau_1 V_1+\tau_2 V_2=0$, shadowing occurs as soon as $V_2<v_b$ , but its effect remains classic as far as $v_b T$ does not become too small compared to the excitation amplitude $Z_{max}=V_1\tau_1/2$ . Everything occurs as if $V_2$ was infinite.

***Classic screening:*** when $V_2<v_b$, there is classic screening; everything occurs as if $V_2$ was infinite and collision occurs only with wall moving at speed $V_1$. Consequently one gets:

$$<v'_b> = v'_b = -\beta v_b + (1+\beta) V_1 \quad (31)$$

Assuming a dissipation $\alpha$ through the gas, the ball comes back with a mean speed $v''_b = -\alpha v'_b$. And self-consistency condition, *i.e.* $\langle v_b \rangle = -\alpha v'_b$, implies:

$$\langle v_b \rangle = -\alpha V_1 (1+\beta)/(1-\alpha\beta) \tag{32}$$

As soon as $\alpha$ is small and $|V_2|>|V_1|$, this approximation is valid except when anomalous screening occurs (see next paragraph) when dissipation in the gas is so large ($\alpha \ll 1$) that $v_b T$ becomes quite small so that the tip of the teeth looks round. However, it can occur also that $\alpha \approx 1$ and $\beta \approx 1$, so that $|\langle v_b \rangle|$ can become large and overpass $|V_2|$. This occurs for the 1-ball case essentially, and this will be treated in the second next paragraph. For instance, considering $\alpha=1$, $\beta=1-\varepsilon=0.05$, the condition for applying Eq. (32) correctly writes $V_2/V_1 > 2/\varepsilon = 40$, which is quite large.

*Anomalous screening:* When $Z_{max} \gg -v_b T$, classic screening ends, because tips of teeth look round at this scale, smoothened by inertia. This can be taken into account by introducing a limit acceleration $\Gamma$, which controls the shape of the tips. This zone lasts a time $\Delta T = \Delta T_1 + \Delta T_2$, such as $\Gamma \Delta T_1 = V_1$ & $\Gamma \Delta T_2 = -V_2$. This leads to similar calculations and trends to those ones developed for sinus-like or parabola-like motions in the same range of conditions. Then identical results will be obtained when written in terms of the acceleration $\Gamma$ as the shadowing.

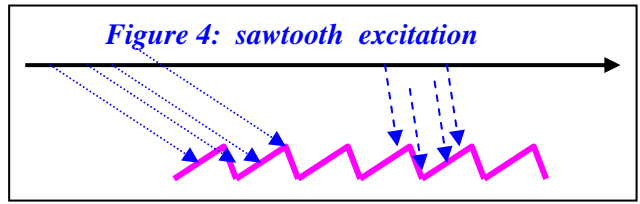

*Figure 4: sawtooth excitation*

*No screening:* This case occurs essentially as soon as the gas does not dissipate ($\alpha=1$) and when the restitution coefficient tends to 1, or when $|V_2| \leq |v_b|$. A ball can collide the piston at speed $V_1$ or $V_2$ with respective probability $p_1=(\tau_1 - 2Z_o/v_b)/T$ and $p_2=(\tau_2+2Z_o/v_b)/T$ and with respective collision rules:

$$v'_{b1} = -\beta v_b + (1+\beta) V_1 \tag{33.a}$$

$$v'_{b2} = -\beta v_b + (1+\beta) V_2 \tag{33.b}$$

so that $\langle v'_b \rangle = p_1 v'_{b1} + p_2 v'_{b2} = -\beta v_b + (1+\beta)[p_1 V_1 + p_2 V_2] = -\beta v_b + (1+\beta)[p_1 V_1 + p_2 V_2] = -\beta v_b + (1+\beta)[\tau_1 V^2_1 + \tau_2 V^2_2]/(T v_b)$, with $v_b<0$, $V_1>0$, $V_2<0$, $2Z_o=V_1\tau_1=-V_2\tau_1$.

$$\langle v'_b \rangle = -\beta v_b + (1+\beta)[2Z_o/(T v_b)] (V_1-V_2) \tag{34.a}$$

Or equivalently:

$$\langle v'_b \rangle = -\beta v_b + (1+\beta)[V_1/v_b] (V_1-V_2) \tag{34.b}$$

Applying self consistency hypothesis, *i.e.* $\langle v''_b \rangle = -\langle v'_b \rangle = \langle v_b \rangle$, leads to the typical mean speed condition $(1-\beta)\langle v_b \rangle = (1+\beta)[V_1/\langle v_b \rangle](V_1-V_2)$; hence to:

$$\langle v_b \rangle = [(1+\beta)/(1-\beta)]^{1/2} [V_1 (V_1-V_2)]^{1/2} \qquad (35)$$

Another time, $\langle v_b \rangle$ tends to infinity when $\beta$ tends toward 1, indicating that the temperature of a ball in a shaken billiard without loss shall be infinite. This result is well-known and obvious because the only way the particle can loose energy in this case is that it meets the boundary when it goes down. Furthermore imposing the same energy transfer for both collision kinds ($V_1$ and $V_2$) imposes vertical trajectories in Fig. (3), hence infinite speed $v_b$.

### *4.d- Conclusion:*

We have seen that anomalous shadowing can take place as soon as losses in the granular gas becomes large ($\alpha \ll 1$). This may inhibit strongly the excitation of the gas, which becomes decoupled from the walls of the container. This effect occurs due to the necessity of overlapping the regions accessible by the ball and by the moving wall. This introduces some correlation in the motion of the boundary, which imposes that ball-wall collisions occur only when the wall is near its maximum position when the ball speed is slower than the typical boundary speed. As the wall position is at maximum, its speed is at minimum, and ball-wall coupling is quite inefficient.

We think that this explains why most experiments on granular gas with cells containing few layers of particles look so quite in 0g when they are excited by sinus shaking. Things are different in 1g because balls are forced to collide with bottom.

To overpass this difficulty, one may try to propose other kinds of mechanical excitation. This is done in the next section.

### **5. Other possible kinds of mechanical excitation:**

In this section we propose few kinds of random excitation which may allow washing out the screening effect due to the correlated motion of boundary.

### *5.a- Case of random excitation with viscous drag:*

One can now consider the case of a random motion produced by some force applied to the system during a series of short lapses of time $\delta t$ and changing randomly of direction (/sign) after each $\delta t$. This produces a random acceleration with a characteristic correlation time $\delta t$. So, this forces the wall speed (see appendix) to perform a random walk, whose mean $\langle V \rangle$ is 0, but whose standard deviation, $\langle V^2 \rangle = (\Gamma^2 T \delta t)$, increases linearly with time. When the system is submitted to some viscous damping, of time constant $\tau \gg \delta t$, then $\langle V^2 \rangle$ saturates at $\langle V^2 \rangle = (\Gamma^2 \tau \delta t)$, and the global system is submitted to a random walk process in the position-space; this X motion obeys a Langevin equation, with diffusive constant $D_X=(\Gamma^2\delta t\ \tau^2)$ (see Appendix).

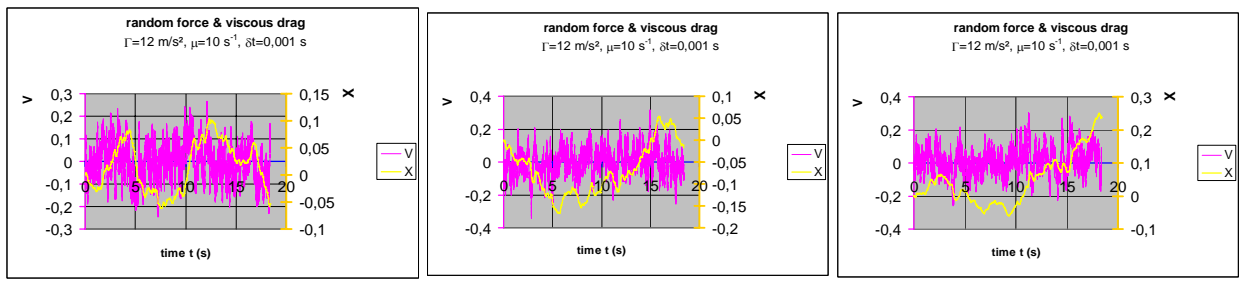

**Figure 5: Simulation of the motion of a wall** submitted to a random force $\pm\Gamma$ with correlation time $\delta t$; the wall obeys the differential equation $d^2X/dt^2= \Gamma(t)-\mu dX/dt$. Mass of the system is taken as unity. 3 realisations at same parameter ($\Gamma= 12$ m/s², $\delta t=0.001$s, $\mu=10$s$^{-1}$). Typical speed is $<|V|> = 0.12$m/s about, as predicted by $<V^2>=(\Gamma^2 \delta t/\mu)$. At a time large compared to $1/\mu$, the X motion becomes a random walk process with a diffusion coefficient $D_x= (\Gamma^2\delta t\ \mu^2)$, predicting a mean square displacement $<X^2>$ on time scale T=20s, which varies as $<X^2>=2D_xT =0.06$m², or $<|X|>$ is 0.25m.

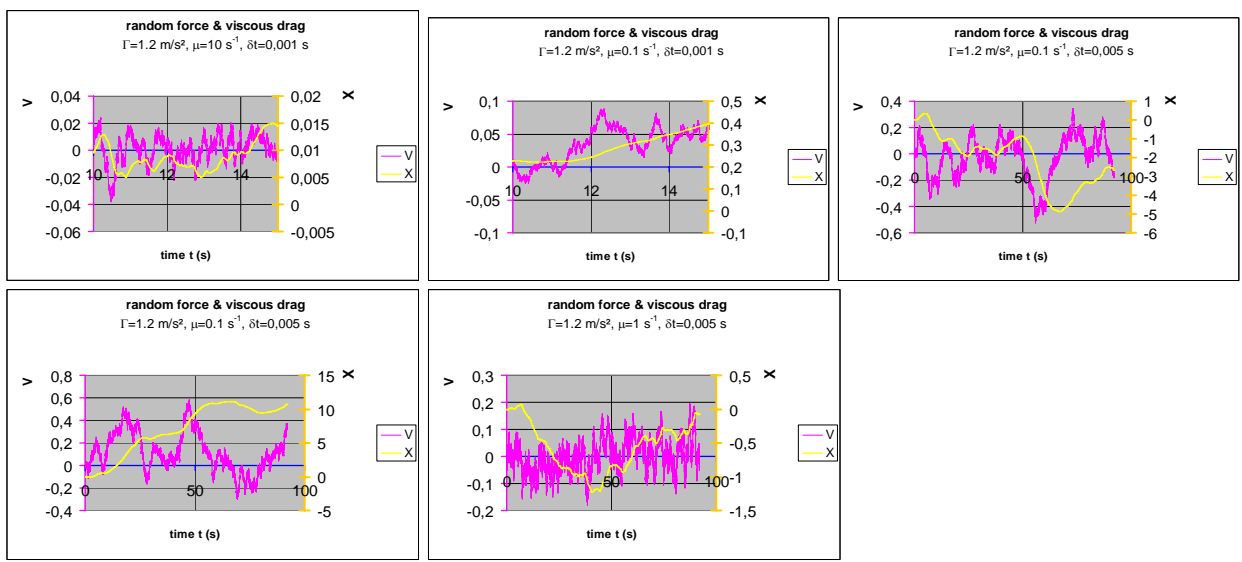

**Figure 6: (same as Fig. 5) : Simulation of the motion of a wall** submitted to a random force $\pm\Gamma$ with correlation time $\delta t$; the wall obeys the differential equation $d^2X/dt^2= \Gamma(t)-\mu dX/dt$. Mass of the system is taken as unity. For different time scales, and values of parameters. One sees that amplitude and speed of the motion can vary strongly with the set of parameters. This method shall require good set of parameters, adapted to the time range of the experiment. One shall note also that each try is different, so that interpreting correctly the data will require the exact knowledge of the force.

$$d^2X/dt^2= \Gamma(t)-\mu(dX/dt) \tag{36}$$

According to the model of random walk, the screening process should be small. A typical example can be the following: Taking $\Gamma=12 m/s^2$, $\delta t=0.001s$, $\tau=0.1s=1/\mu$, one gets $<V^2>=(\Gamma^2 \delta t/\mu^2)=144*0.1*0.001=0.0144 m^2/s^2$, while $D_X= 144*0.001*0.1*0.1=0.00144 m^2/s$. Few tries with this set of parameters are displayed in Fig. 5. Simulations with other sets of parameters are displaid in Fig.6.

### *5.b- Case of random excitation with elastic constraint:*

The problem generated by the previous modelling is that the system is predicted to perform a random walk at long time, so that the standard deviation $<X^2>$ of the position of its centre of mass increases linearly with time. This is not allowed in real micro-gravity condition. So, we have simulated the motion of a system with unit mass excited by a random force $\pm\Gamma$ with typical correlation time $\delta t$ and submitted to an elastic force with spring stress constant k. The system obeys Eq. (37). Results of simulations are reported in Fig 7.

$$d^2X/dt^2 = \Gamma(t) - kX \qquad (37)$$

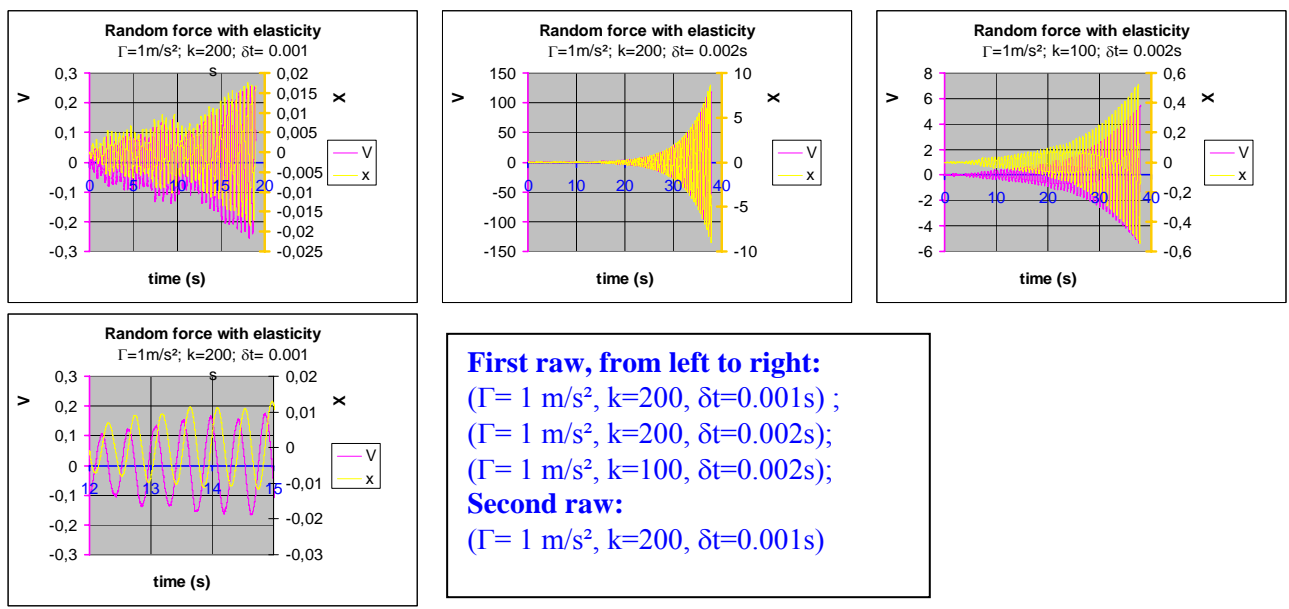

**Figure 7: Simulation of the motion of a wall** submitted to an elastic link (spring strength constant k) and to a random excitation $\pm\Gamma$ with correlation $\delta t$; the wall obeys the differential equation $d^2X/dt^2= \Gamma(t) -kX$. Mass of the system is taken as unity. First line: from left to right ($\Gamma= 1$ m/s², k=200, $\delta t=0.001s$) ; ($\Gamma= 1$ m/s², k=200, $\delta t=0.002s$); ($\Gamma= 1$ m/s², k=100, $\delta t=0.002s$). One sees an oscillation at $\omega^2=k$, whose amplitude diverges after a while because of the absence of damping. Frequency of oscillation can be checked on Fig. (d/bottom-left) on the second line, which corresponds to the ($\Gamma= 1$ m/s², k=200, $\delta t=0.001s$) case with a dilation of scale; one gets $f_{resonance}=\omega/(2\pi)\approx 2.25$ Hz.

We see first that the mean position remains located at 0. We see also that the system is able to store some kinetic energy on a single mode, which corresponds to the resonance: ω²=k. Hence the amplitude of oscillation at this frequency becomes extremely large after a while. This is not allowed in micro-gravity experiment. Furthermore, since the energy stored in this mode is large, the motion starts being sinus shaped and gets the disadvantage of shadowing of previous excitation.

Anyhow, this kind of motion is not possible, since it exists always some viscous damping.

Beside, we have tested the effect of a biased motion provoked by a random force ±Γ with a not zero mean, <Γ> ≠ 0. In this case the mean position is displaced at <X>= - <Γ> /k.

## 5.c- *Case of random excitation with elastic constraint and viscous damping:*

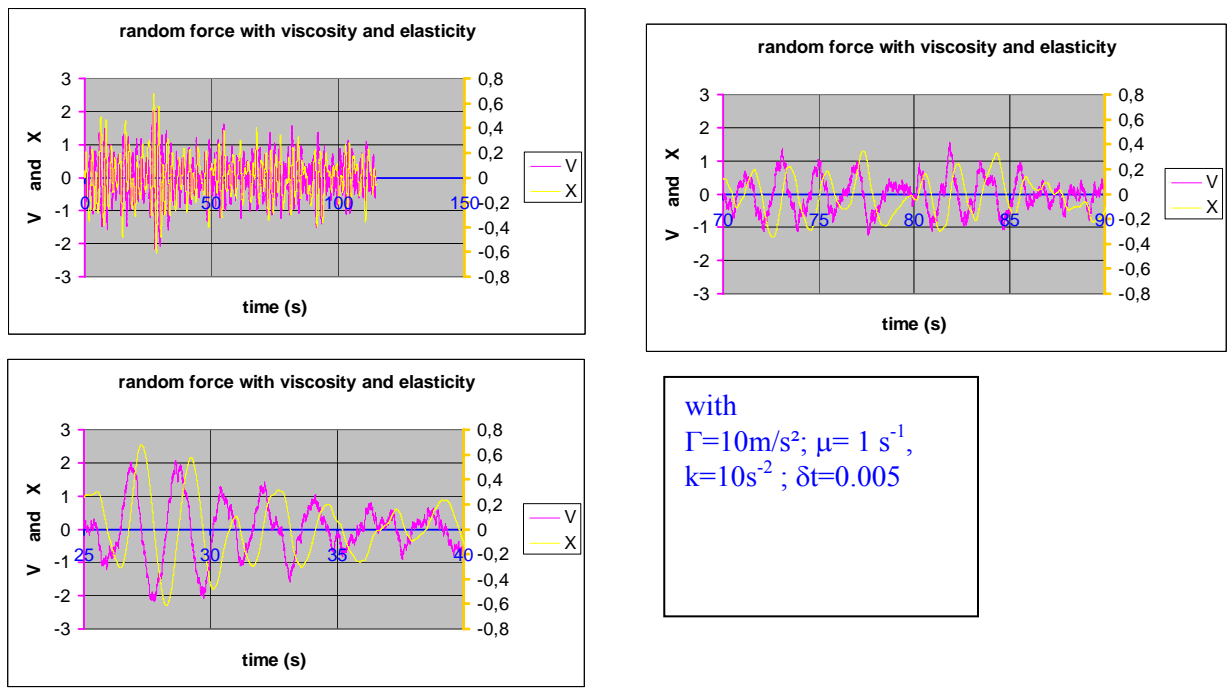

**Figure 8: Simulation of the motion of a wall** submitted to a viscous damping (-μv) and an elastic link (spring strength constant k) and a random excitation ±Γ with correlation time δt; the wall obeys the differential equation d²X/dt²= Γ(t)−μ(dX/dt) −kX.
The 3 Figures have been obtained with Γ=10m/s², μ= 1 s⁻¹, k=10s⁻² and a step of time δt=0.005, at different time scale. One sees that collisions with boundary can arrive with fast boundary; but the collisions are rare. Oscillations are induced by resonance at frequency f=ω/(2π) such as ω²=k=>f ≈0.5Hz. <V²>=Γδt/μ=0.05.

## 5.c- *Case of random excitation with elastic constraint and viscous damping:*

To forbid the storage of energy on the oscillating resonant mode, one can damp the system. This is done in this subsection. So, we have simulated the motion of a system with unit mass, excited by a random force ±Γ with typical correlation time δt and submitted to a viscous force of viscous coefficient μ and to an elastic force with spring stress constant k. Eq. (38) is the governing equation. Results of simulations are reported in Figs 8-9-10.

$$d^2X/dt^2 = \Gamma(t) - \mu(dX/dt) - kX \tag{38}$$

Simulations show that one can get efficient collisions of balls which are merely at rest with rapid boundaries, but these are rare.

We have checked that a biased motion, provoked by a random force ±Γ with a non zero mean, $\langle\Gamma\rangle \neq 0$ does not disturb the behaviour. It displaces the mean position X at $\langle X\rangle = -\langle\Gamma\rangle/k$.

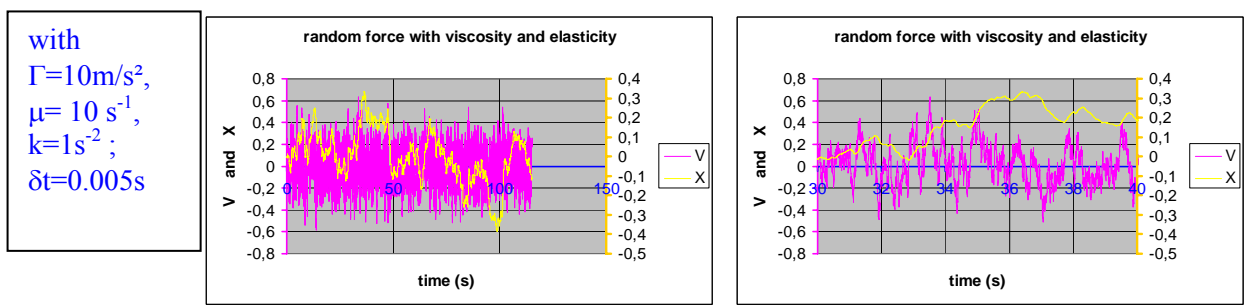

**Figure 9:** Same as Figure 8, except that Γ=10m/s², μ= 10 s⁻¹, k=1s⁻² ; δt=0.005s. (The wall obeys the differential equation d²X/dt²= Γ(t)−μ(dX/dt) −kX).

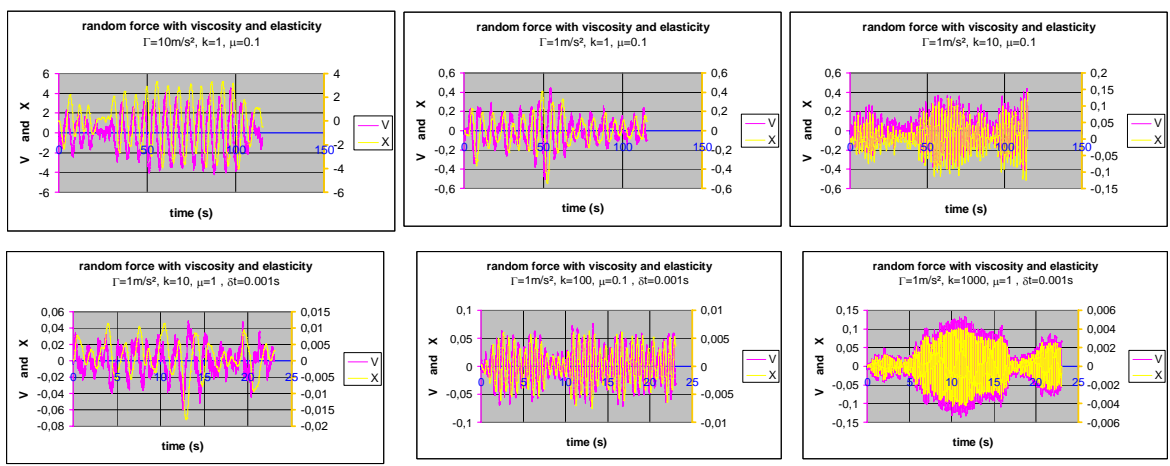

**Figure 10:** Same as Figure 8, (System obeys the differential equation d²X/dt²= Γ(t)−μ(dX/dt) −kX). The different sets of values are: first line: δt=0.005s and from left to right: (Γ=10m/s², μ=0.1 s⁻¹, k=1 s⁻²); (Γ=1m/s², μ=0.1 s⁻¹, k=1 s⁻²); (Γ=1m/s², μ=0.1 s⁻¹, k=10 s⁻²); δt=0.005s. Second line: δt=0.001s and from left to right: (Γ=1m/s², μ=1 s⁻¹, k=10 s⁻²); (Γ=1m/s², μ=1 s⁻¹, k=100 s⁻²); (Γ=1m/s², μ=1 s⁻¹, k=1000 s⁻²)

# 6. Discussion and conclusion

This paper has studied the conditions of excitation of a dissipative granular gas in micro-gravity excited by moving boundaries. It has emphasized the importance to take into account in detail the real motion to be able to predict the coupling. Few results have been obtained:

1) Through a detailed analysis of the trajectories of both the moving walls and the balls, the paper investigates the effect of periodic forcing, and demonstrates that a gas of particles which is strongly dissipative is badly coupled to the moving boundaries. This is induced by a shadowing effect which "hides" the boundary to slow ball during the part of the period when the wall moves fast.
2) According to this effect, this paper demonstrates that the approximation which consists in calculating the time at which the ball-wall collision occurs as if the wall was immobile or in its mean position is strongly incorrect since one deduces directly from this time (i) a wrong phase of the vibration and (ii) a wrong speed of the wall during the collision. This "solution" allows collision with the wall when it moves fast, which is not allowed as soon as the dissipation in the gas is important and the ball speed is small. This is worth to be emphasised, since this approximation is used in many computer simulations. However, the method can become correct if one considers the time t' at which the ball crosses the maximum position of the piston and calculate t from t', as we have done (Fig. 3).
3) It results that slow balls are poorly coupled to boundaries moving periodically.
4) A mean field argument with a self-consistent condition has been developed to predict the typical particle speed. For instance it is found that the coupling with boundary of a strongly dissipative gas is so bad that the scaling law which fixes the variation of its typical speed $<v_b>$ with the dissipation is modified importantly. However, one finds also that this typical speed $<v_b>$ scales always linearly with the boundary typical speed $<V>$.
5) The paper analyses the difference generated by different types of boundary motion (sinus, arches of parabola, sawtooth, random motion with viscous drag).
6) An other consequence is the following: Since the excitation efficiency depends on the particle speed, it could happen that the particles of the dissipative gas "separates" spontaneously into two classes when the mean free path of the particle is of the order of a part of the cell size L: the class 1 will be made of slow particles which are mainly uncoupled to boundaries, and cool down by collisions the energy of the fastest particles; and class 2 is formed by particles which are the fastest and interact strongly with the moving boundaries, gaining energy at each collision with boundary. These two classes exchange particles at random by mutual collision. To be observable, this regime requires that the probability that a fast particle crosses the cell without colliding a slow particle shall be not too small. Hence it requires that the cell size L is a few times only the mean free path $l_c$ of the particles. Hence, this paper strengthens the model of biphasic granular gas [9-10, 12] which we have recently proposed to explain the experimental distribution of ball speed observed at intermediate densities [13, 14], *i.e.* $\rho(v_b)=(1/v) \exp\{-v_b/v_o\}$.

7) The paper demonstrates also that the correct macroscopic conditions shall at least take into account the number of grains and some "surface-to-volume" ratio to take into account correct dissipation for computing the typical speed and the mean condition of excitation. This boundary condition cannot be summed up by giving simply a constant temperature or a constant typical speed. Such a conclusion has been drawn from some other works in some other context [19].

8) The nature of the difficulty can be emphasized already by the study of the dynamics of a single ball as a function of its restitution coefficient $\beta$. When $\beta$ is 1 the mean speed of the ball shall diverge since the ball shall gain in average as energy as it shall loose during the collision with the walls. This forces $<v_b>/<V>$ to diverge. When $\beta$ is slightly smaller than 1, an equilibrium is reached which fixes the ratio $<v_b>/<V>$ to values larger than 1. The ball speed can become much smaller than $<V>$ when $\beta$ becomes much smaller than 1, and this generates the shadowing, an increase of the reduction of the speed, and a change of scaling when $\beta<<1$ (Eqs.19 & 24).

9) The effect of gravity shall be mentioned at this stage: as a matter of fact, the above analysis is only valid when no gravity drives the balls towards the moving wall. Indeed, the dynamics of a completely inelastic ball will be not at all the same [20], for which ball speed can be large, even when $\beta=0$.

10) At last, it is worth noting that some recent experiments performed in a Chinese satellite (SJ-8) in collaboration with Prof. Hou tend to demonstrate that dense vibrated cells behave in such a way that one observes the grains localised as some dense zone of granular matter in the middle of the cell, which is mainly at rest; the grains seems not moving in the satellite frame, except during some rare lapses of time during which some of the grains moves rather slowly independently from the others before stopping, or during which the zone rotates or translates coherently changing of shape when it meets the lateral boundaries [21]. They question where is the gas. Are these preliminary experimental results an evidence of what we are saying? It seems that the present paper may explain them qualitatively.

11) In fact, in these last experiments [21], we have found also in an other cell, which is merely 2-d and filled with much less grains, that the speed distribution $\rho_{2-d}(v_b)$ of the particle speed at low density varies as $\rho_{2-d} =\exp(-v_b/v_o)$. It is important to note that this result is not similar to the trend obtained in 3-d samples [13-14,8-9], *i.e.* $\rho_{3-d}(v_b)=(1/v) \exp\{-v_b/v_o\}$, since this one exhibits a divergence when v tends to 0, which forces us to claim to the biphasic nature of the 3-d gas. On the contrary, the behaviour $\rho_{2-d}(v_b)= \exp\{-v_b/v_o\}$ does not diverge at $v_b \rightarrow 0$; this last trend is linked to dissipation due to friction on lateral walls most likely.

From this work, one may conclude that it is still surprising that no simulation has been focusing on the problem of the coupling between balls and boundaries and has treated extensively on the difficulty of exciting correctly a gas of dissipative particles, as soon as they collide efficiently. This should not be difficult to obtain with only few hundreds of particles.

Turning now to the problem of finding and observing the gas-cluster transition, which should occur at rather large concentration, we shall emphasize that the excitation of the dissipative gas in this gas-cluster transition may be quite difficult with periodic vibration due to the very poor coupling of the slow particles to the boundaries. In fact if this transition exists and the separation in two phases obtained, this would tell that the gas which is observed is in the Knudsen regime most likely.

## Appendix:

**Piston with random motion:**

We consider a piston moving according to a random walk obeying a Langevin equation [21]

$M_p dV/dt = F(t) - M_p \mu V$     or     $dV/dt = \Gamma(t) - \mu V$

With V the piston speed, $M_p$ the piston mass, $M_p\mu$ the viscous friction coefficient and $F(t)=M_p\Gamma(t)$ a random force with zero time-mean. $\Gamma(t)$ is some random function with a typical correlation time $\tau_c$ small compared to $1/\mu$, so that $<\Gamma(t)\Gamma(t+\tau)>=g(\tau)$ is non zero when $\tau$ is small; we define D as $\int_0^\infty g(\tau) d\tau = D$.

If $\Gamma$ is constant during $\delta t$, then $D = \Gamma^2 \delta t$

Starting from a speed $V_o$, the average speed of the piston obeys $<V_p>=V_o \exp(-\mu t)$ which tends to 0 within a time scale $T_R=1/\mu$; whereas the speed standard deviation $\sigma^2(t)=<V(t)^2-<V(t)>^2>$ is given by

$\sigma^2(t) = 2D\int_0^\infty \exp(-2\mu\{t-t'\})dt' = \{D/\mu\}[1-\exp(-2\mu t)]$

Hence at short time t the speed $V_p$ diffuses and its standard deviation increases linearly with time as $\sigma^2 = 2Dt$, then it saturates and remain constant after a time $t>T_R=1/\mu$ with $\sigma^2=D/\mu$.

([21] N. Pottier, cours de DEA (1997), Paris, p.305-310):

Now if we consider the position Z of the piston, it obeys $d^2Z/dt^2 = \Gamma(t) - \mu \, dZ/dt$. This motion tends to a diffusion at long time (*i.e.* $t>T_R=1/\mu$), *cf.* N. Pottier p.308) with a diffusion coefficient $D_z=(D/\mu^2)$ so that $<Z^2(t)> \to 2D_z t = 2(D/\mu^2) t$ at time $t>>T_R$.

*Acknowledgements:* CNES and ESA are thanked for funding, CNSA is gratefully thanked. This is part of a micro-gravity program on the study of the behaviours of granular matter in 0g conditions with main collaborations: M. Leconte, F. Douit, M. Hou, Y. Garrabos, C. Lecoutre, F. Palencia, D. Beysens, S. Fauve, E. Falcon, who are gratefully thanked.